\documentclass[11pt]{article}
\usepackage[final]{acl}
\usepackage{times}
\usepackage{latexsym}
\usepackage[T1]{fontenc}
\usepackage[utf8]{inputenc}
\usepackage{microtype}
\usepackage{inconsolata}
\usepackage{graphicx}
\usepackage{booktabs}
\usepackage{amsmath,amssymb}
\usepackage{xcolor}
\usepackage{multirow}
\usepackage{placeins}
\usepackage{enumitem}
\usepackage{colortbl}
\usepackage{xspace} % Optional, helps handle spacing automatically
\definecolor{mpos}{HTML}{649EE5}
\definecolor{mneg}{HTML}{EA7F7C}
\newcommand{\rR}{\textsf{R}}
\newcommand{\rD}{\textsf{D}}
\newcommand{\rQ}{\textsf{Q}}
\newcommand{\rA}{\textsf{A}}

\newcommand{\ourmethod}{\textsc{CO-GRAFT}\xspace} 

\title{\ourmethod: Attributing Reinforcement Learning Gains in Search Agents via Role-Level Policy Grafting}

\author{
  Ramraj Chandradevan, 
   Sayontan Ghosh, Vinoth Selvendran \\
  \texttt{\{first.last\}@gmail.com}
}

\begin{document}
\maketitle

\begin{abstract}
Reinforcement learning (RL) improves large language model search agents by up to 32 exact-match (EM) points over their pre-RL initializations, yet overall accuracy gains do not identify which underlying functional capabilities improve or interact during training. We introduce \ourmethod, a reward-agnostic interventional attribution framework in which an RL-trained agent and its pre-RL initialization jointly generate a single trajectory under closed-loop execution with live retrieval. Each policy controls a mutually exclusive subset of four functional roles: intermediate reasoning, search-or-answer decision, search query generation, and final answer extraction. Evaluating all $2^4=16$ role assignments without modifying weights or retraining defines a cooperative game whose Shapley values decompose the total RL gain across roles, while M\"obius interaction terms distinguish standalone role improvements from multi-role coupling. Applying \ourmethod to Search-R1 agents across Qwen2.5 and Llama-3.2 models (spanning two sizes, base and instruction-tuned initializations, GRPO and PPO optimizers, and NQ and HotpotQA datasets), we show that query generation never exceeds 15\% of the total gain; instead, gains are governed by pre-RL answer validity, search retention, model scale, and fixed task-level search budgets, shifting credit from reasoning toward answer extraction in larger or well-formatted answer response checkpoints while smaller base models rely on reasoning--decision coupling.
% we demonstrate that RL gains are driven primarily by pre-RL answer validity, search retention, model scale, and task-level search budgets rather than query formulation or optimizer choice.
\end{abstract}

\section{Introduction}
\label{sec:intro}

% 1. Task: Defining and explaining LLM multi-turn search agents
Large language model (LLM) search agents answer multi-hop and knowledge-intensive questions by interleaving reasoning with external retrieval across multiple turns \cite{jin2025searchr1,song2025r1searcher,chen2025research,zheng2025deepresearcher,sun2025zerosearch,gao2025asearcher,guo2025deepseekr1}. At each turn, the agent executes four distinct functional roles in sequence \cite{jin2025searchr1,song2025r1searcher,chen2025research}: it generates intermediate reasoning (\rR), makes a binary decision (\rD) either to invoke search or to terminate and answer, formulates a search query (\rQ) when search is chosen, and extracts the final answer (\rA) once retrieval concludes. Post-training these agents with reinforcement learning (RL) improves exact-match (EM) accuracy by 6 to 25 points over non-RL inference and retrieval baselines \cite{jin2025searchr1,jin2025empirical}, and by up to 32 points over prompted pre-RL search agents in our evaluation.

% 2. Problem: Final EM gains obscure which functional capabilities improve under any RL reward scheme
However, overall EM improvements do not reveal which functional capabilities change during RL post-training or how multiple capabilities interact to increase accuracy \cite{jin2025empirical}. Whether an agent is trained with outcome rewards \cite{jin2025searchr1,chen2025research,guo2025deepseekr1}, process rewards \cite{lightman2024lets,wang2024math,zheng2025deepresearcher}, or auxiliary reward shaping such as format \cite{jin2025searchr1,guo2025deepseekr1} and retrieval-utility rewards \cite{song2025r1searcher,sun2025zerosearch,gao2025asearcher}, an increase in final EM can result from four distinct role behaviors \cite{jin2025searchr1,jin2025empirical}: formulating more effective search queries, reasoning more accurately over retrieved passages, deciding more appropriately when to search versus answer, or formatting final answers to match evaluation criteria. Identifying which of these roles actually drives EM gains is critical, especially for outcome-reward RL \cite{jin2025searchr1,guo2025deepseekr1}, where supervision is applied only to the final answer, yet applies generally to any trained search agent for diagnosing failure modes and targeting future training improvements.

% 3. Limitations of prior work: Observational metrics, single-role ablations, offline prefixes, and training-time credit assignment
Existing evaluation and credit-assignment approaches do not isolate the causal contribution of each functional role under closed-loop execution. Observational behavioral analyses measure surface statistics, such as response length \cite{jin2025searchr1,guo2025deepseekr1}, search frequency \cite{jin2025searchr1,jin2025empirical}, or query lexical overlap \cite{jin2025empirical}, which correlate with training progress without verifying whether changes in a given role actually increase end-to-end EM accuracy. Single-role ablations \cite{jin2025empirical} perturb one component in isolation and therefore miss multi-role interactions where two or more roles must be controlled jointly by the RL policy to produce a gain. Offline trajectory interventions \cite{zheng2025deepresearcher} evaluate static prefixes rather than closed-loop execution (interactive feedback with live retrieval), whereas process-reward \cite{lightman2024lets,wang2024math} and step-level credit-assignment methods \cite{gao2025asearcher} alter reward signals during optimization rather than decomposing the learned capabilities of an already-trained checkpoint.

\begin{figure*}[t]
\centering
\includegraphics[width=\textwidth]{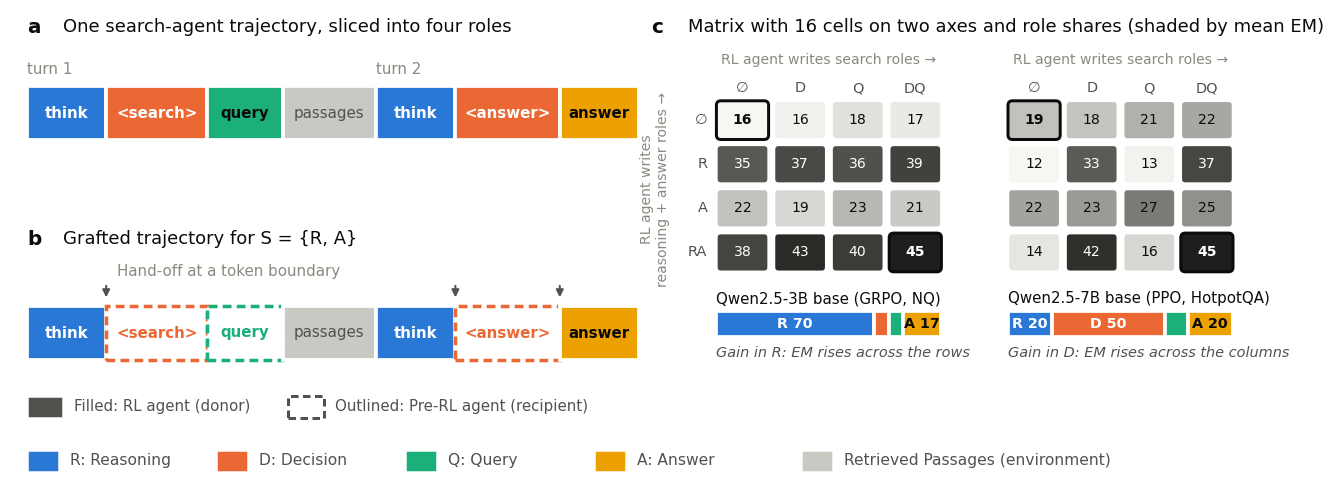}
\caption{Overview of the \ourmethod framework for attributing RL performance gains. (a) A search-agent trajectory partitioned into four roles: reasoning \rR, decision \rD, query \rQ, and answer \rA; gray passages are returned by the environment. (b) A grafted trajectory for $S=\{\rR,\rA\}$: the RL agent (filled) generates \rR\ and \rA, while its pre-RL model (outlined) generates \rD\ and \rQ\ during live retrieval. (c) The $4 \times 4$ grid of all 16 role combinations, indexed by the roles owned by the RL agent ($\text{row} \cup \text{column}$) while the pre-RL model generates the remaining roles: columns add search roles (\rD, \rQ) and rows add reasoning and answer roles (\rR, \rA), progressing from $\emptyset$ (top left, pre-RL model alone) to $\{\rR,\rD,\rQ,\rA\}$ (bottom right, RL agent alone). Each cell reports the mean EM (\%) of the grafted policy on the same 1,000 evaluation questions (darker shading indicates higher EM), alongside each role's Shapley percentage share of total gain $v(N)-v(\emptyset)$ for Qwen2.5-3B base (GRPO, NQ) and Qwen2.5-7B base (PPO, HotpotQA).}

\label{fig:mechanism}
\end{figure*}

% 4. Method overview: CO-GRAFT (Co-Policy Grafting), 16 role assignments, Mobius terms, and Shapley values
To overcome these limitations, we introduce \textbf{\ourmethod} (\textbf{Co-Policy Grafting}), a reward-agnostic interventional attribution framework in which an RL-trained agent and its pre-RL initialization jointly generate a single trajectory with live retrieval. Each policy controls a mutually exclusive subset of the four roles within the same trajectory (Figure~\ref{fig:mechanism}b), generating the token spans for its assigned roles and passing the resulting prefix to the other policy at role boundaries. Evaluating all $2^4 = 16$ role combinations defines Shapley~\cite{shapley1953} values without modifying model weights, requiring access to training rewards, or retraining. From these 16 measurements, M\"obius interaction terms \cite{harsanyi1963} separate standalone single-role improvement from pairwise and higher-order (multi-role) dependencies, and Shapley values aggregate each role's marginal contribution across all role combinations into a single additive attribution score.

% 5. Key findings: Concrete empirical takeaways across models and datasets
Applying \ourmethod to Search-R1 agents \cite{jin2025searchr1,jin2025empirical} across Qwen2.5 \cite{qwen2.5} and Llama-3.2 \cite{llama3} models (spanning two model sizes, base and instruction-tuned initializations, and GRPO \cite{shao2024deepseekmath} and PPO \cite{schulman2017proximal} optimizers on NQ \cite{kwiatkowski2019nq} and HotpotQA \cite{yang2018hotpotqa} datasets) reveals how RL redistributes capability across roles. First, search query generation (\rQ) never accounts for more than 15\% of the total RL gain on either dataset, showing that improved retrieval queries are not the primary driver of accuracy gains. Second, pre-RL answer validity, search retention, and model scale govern how credit is distributed between reasoning and answer extraction: larger models and instruction-tuned checkpoints that already produce well-formed answers before RL shift credit from reasoning (\rR) toward final answer extraction (\rA), whereas smaller base models exhibit strong positive coupling between reasoning and decision (\rR$\times$\rD). Third, decision (\rD) gains reflect learning a fixed task-level search budget rather than deciding when to stop based on question-level evidence sufficiency.

% 6. Contributions: Concise, non-repetitive bullet list
We summarize our contributions as follows:
\begin{itemize}
    \item We propose \textbf{\ourmethod}, an interventional attribution framework that segments a search agent trajectory into four functional roles (\rR, \rD, \rQ, and \rA) and evaluates all $2^4 = 16$ role-assignment combinations between an RL-trained agent and its pre-RL initialization under live retrieval without modifying weights or retraining.
    \item We formulate an exact decomposition of RL total EM gains via Shapley values and M\"obius interaction terms, isolating standalone role improvements from pairwise and higher-order interactions.
    \item We evaluate Search-R1 agents using our framework across Qwen2.5 and Llama-3.2 models on NQ and HotpotQA, showing that RL gains are driven by pre-RL answer validity, search retention, model scale, and fixed task-level search budgets while query generation remains bounded at $\le 15\%$ of total gain.
\end{itemize}

\section{\ourmethod}
\label{sec:method}

We introduce \ourmethod, an attribution framework that decomposes the performance gain of an RL-trained search agent across the functional roles of its trajectory. To isolate where RL improvements originate, we first partition every trajectory into four exhaustive, non-overlapping roles. We then construct grafted rollouts in which the RL policy and its pre-RL initialization jointly generate trajectories under controlled role assignments, and finally quantify both standalone and synergistic contributions using Shapley decomposition and M\"obius interaction terms.

\paragraph{Roles.}
\ourmethod decomposes a search-agent trajectory into four distinct roles covering every policy-generated token. Following the standard Search-R1 trajectory format \cite{jin2025searchr1}, an agent interleaves reasoning blocks (\texttt{<think>}$\,r\,$\texttt{</think>}) with search calls (\texttt{<search>}$\,q\,$\texttt{</search>})—each followed by environment-returned passages enclosed in \texttt{<information>}\ldots\texttt{</information>} tags—before terminating with a final response (\texttt{<answer>}$\,a\,$\texttt{</answer>}; Figure~\ref{fig:mechanism}a). We formalize four \emph{functional roles}, $N=\{\rR,\rD,\rQ,\rA\}$: \rR\ encompasses all reasoning tokens generated prior to the next action tag; \rD\ governs the discrete action selection (\texttt{<search>}, \texttt{<answer>}, or an incomplete-turn termination); \rQ\ comprises the search query string; and \rA\ contains the final answer text. Because \texttt{<information>} blocks are injected externally by the retriever rather than sampled from either policy model, they belong to no player. Crucially, this partition assigns every model-generated token in the trajectory to exactly one role, ensuring that any shift in task accuracy traces directly to the specific roles whose output tokens change.

\paragraph{Grafted rollouts.}
A grafted rollout is a trajectory generated jointly by the RL agent and its pre-RL model, each owning a non-overlapping subset of the roles (Figure~\ref{fig:mechanism}b). Let $\pi^{+}$ denote the RL-trained agent (the \emph{donor}) and $\pi^{-}$ denote its exact pre-RL initialization (the \emph{recipient}), both sharing the same tokenizer and prompt template. For a grafted policy $S\subseteq N$, every role segment $\rho$ is generated by $\pi^{+}$ if $\rho\in S$ and by $\pi^{-}$ otherwise. During generation, whichever policy writes the active segment conditions on the full preceding prefix, including tokens produced by the paired model and live passages returned by the retriever for each search query. To prevent artifacts during policy hand-offs, our harness enforces three invariance constraints: (1) because a model generates reasoning and its subsequent action tag in one pass, the harness discards the tag produced by the owner of \rR\ and lets the owner of \rD\ decode the action tag from the identical prefix; (2) all hand-offs occur strictly at whole-token boundaries so split tokens (e.g., ``\texttt{\textvisiblespace<}'' in `` \texttt{<search>}'') do not distort downstream tokenization; and (3) the harness parses an action only when a complete opening-and-closing tag pair is generated, treating malformed or unclosed tags as invalid turns that pass an error prompt to the next step. All remaining execution mechanics follow the standard search-agent inference loop, recovering pure recipient behavior at $S=\emptyset$ and pure donor behavior at $S=N$.

\paragraph{Shapley attribution.}
Evaluating all $2^{4}=16$ role combinations (Figure~\ref{fig:mechanism}c) defines a cooperative game whose Shapley values provide an exact, order-invariant attribution of the RL gain. Let $v(S)$ denote the mean Exact Match (EM) score of grafted policy $S$ across a fixed evaluation set, with $v(\emptyset)$ and $v(N)$ representing the recipient and donor baselines, respectively. The Shapley value of role $i\in N$ \cite{shapley1953} is
\begin{equation}
\phi_i=\sum_{S\subseteq N\setminus\{i\}} w_{S}\,\bigl[v(S\cup\{i\})-v(S)\bigr],
\label{eq:shapley}
\end{equation}
where $w_{S}=|S|!\,(|N|-|S|-1)!/|N|!$ weights each marginal contribution by the fraction of the $4!$ role permutations in which the subset $S$ precedes role $i$. By the efficiency axiom, $\sum_{i\in N}\phi_i=v(N)-v(\emptyset)$ holds identically, guaranteeing that the four role attributions sum to the exact total RL gain. We report role \emph{shares} as normalized percentages (the bars in Figure~\ref{fig:mechanism}c) of this total gain ($\phi_i / [v(N)-v(\emptyset)]$). Because percentage normalization becomes unstable near zero gain, we report shares only when the overall RL gain reaches at least four EM points (four times the single-point decoding noise floor), labeling smaller differences as ``no gain.''

\paragraph{M\"obius interactions.}
While Shapley values aggregate each role's total credit into a single scalar, M\"obius interaction terms \cite{harsanyi1963} split that credit into what a role gains alone versus what it gains only together with other roles. For every role combination $T\subseteq N$ of two or more roles ($|T|\ge 2$), the interaction term is
\begin{equation}
m(T)=\sum_{S\subseteq T}(-1)^{|T|-|S|}\,v(S).
\label{eq:moebius}
\end{equation}
Intuitively, $m(T)$ isolates the residual gain realized only when $\pi^{+}$ simultaneously controls all roles in $T$: it becomes zero when roles contribute independently, turns positive under synergistic coupling, and turns negative when roles overlap as substitutes. Because raw interaction magnitudes remain informative even when overall RL gains are small, we report M\"obius terms directly in EM points across all evaluated models.

\section{Experimental Setup}
\label{sec:setup}

In this section, we detail our search agents, datasets, metrics, training recipe, retrieval setup, trajectory harness, and auxiliary control tests to study the RL performance gain attribution across agent roles using our proposed \ourmethod framework. We examine the publicly available Search-R1~\cite{jin2025searchr1,jin2025empirical} checkpoints, and supplement them by training additional models using the same Search-R1 training and evaluation setting.

\subsection{Agents}
\label{sec:agents}

\paragraph{Recipients.} We choose search agents across two families $\times$ two starting points $\times$ two sizes. The first family comprises Qwen2.5-3B and Qwen2.5-7B, each as \emph{base} and \emph{instruct} \cite{qwen2.5}. The second comprises Llama-3.2-1B and Llama-3.2-3B, each as \emph{base} and \emph{instruct} \cite{llama3}. Every recipient is run under its donor's training prompt, so $v(\emptyset)$ is the prompted pre-RL model that prior work reports as the baseline \cite{jin2025searchr1}.

\paragraph{Donors.} For each corresponding recipient model above, we obtain an RL-trained (GRPO) donor agent. For Qwen2.5-3B, Qwen2.5-7B and Llama-3.2-3B (base and instruct) the donors are available publicly. For Llama-3.2-1B (base and instruct), we train the donors ourselves by following the same recipe (\S\ref{sec:recipe}).

\subsection{Evaluation Datasets and Metrics}
\label{sec:data}
We use the same evaluation datasets from the Search-R1 setting.
Because running every question through 16 role assignments is computationally expensive, we evaluate on fixed random subsets rather than full test sets: 1,000 questions each of NQ \cite{kwiatkowski2019nq} and HotpotQA \cite{yang2018hotpotqa}. 
Our primary metric is EM on the last answer block, and we also report the valid-answer rate (the fraction of questions with a well-formed answer block).

\subsection{Training Recipe}
\label{sec:recipe}
We adapt the Search-R1 GRPO recipe to train search agents, including outcome EM reward on the final answer plus a 0.2 format bonus for a well-formed trajectory, retrieved tokens masked from the loss, four search turns of at most 500 generated tokens per turn, group size 5, learning rate $10^{-6}$, batch size 512, KL coefficient $10^{-3}$, and 1$k$ steps of 240 prompts. Moreover, we train on the combined training split of NQ and HotpotQA and evaluate separately on both test sets.

\subsection{Retrieval Environment}
\label{sec:retrieval}
Following the Search-R1 setup \cite{jin2025searchr1}, we use E5-base-v2 \cite{wang2022e5} as the dense dual-encoder over the 21M-passage 2018 Wikipedia dump with exact inner-product search. The retriever is queried live during every rollout, returning the top three passages as retrieved evidence.

\subsection{Harness}
\label{sec:impl}
To execute multi-turn rollouts, each trajectory in both training and evaluation is bounded to at most four search turns followed by one final answer turn, with a maximum of 500 generated tokens per turn. Any turn that terminates without a complete action tag receives an invalid-action prompt and passes to the next turn if the turn budget permits.

Because trajectory grafting modifies rollout execution flow, we verify that the \ourmethod harness preserves single-policy behavior when no role switching occurs, by evaluating $S=\emptyset$ (pre-RL agent only) and $S=N$ (RL agent only), under the original Search-R1 codebase and the \ourmethod harness across identical evaluation sets. Across 32 comparisons (8 model pairs, two datasets, and two policy behaviors), \ourmethod matches Search-R1 accuracy within single-point EM noise in 28 cases, with the remaining four cases differing by at most 2.3 points. Thus, the harness reproduces baseline policy behaviors without introducing implementation artifacts.

\subsection{Control Experiments}
\label{sec:tests}
We run three control experiments to isolate role attributions from formatting artifacts, fixed search budgets, and query-text conditioning.

\paragraph{Answer Content vs.\ Formatting.}
\label{sec:format}
To separate tag compliance from factual gains under strict EM, we recompute attributions using lenient metrics (substring match, token F1, and valid-answer rate). Under lenient scoring on NQ, \rA\ attribution collapses for Qwen2.5 models (e.g., from 38\% to $-5$\% on Qwen2.5-7B base) but persists for Llama-3.2-1B-Instruct (12 vs.\ 16 EM points), showing that \rA\ gains reflect format compliance in Qwen2.5 and content improvement in Llama-3.2. However, Table~\ref{tab:grid_main} reports strict EM.

\paragraph{Adaptive Stopping vs.\ Fixed Search Depth.}
\label{sec:budget}
To test whether \rD\ learns per-question adaptive stopping or a uniform search count, we replace \rD\ with a fixed budget $B \in \{0, \dots, 4\}$ while retaining the RL agent's remaining roles. Across all evaluated pairs, the optimal fixed budget matches the full RL agent within one EM point (e.g., $B=1$ on NQ and $B=2$ on HotpotQA for Qwen2.5-3B PPO), indicating a learned task-level search depth rather than adaptive stopping.

\paragraph{Passage Retrieval vs.\ Query Conditioning.}
\label{sec:swap}
To separate retrieval quality from query-text conditioning, we hold the RL agent's retrieved passages fixed across all 16 combinations on Qwen2.5-3B and Qwen2.5-7B (PPO). Freezing passages drops \rQ\ attribution by 60--100\% (0.6--2.5 EM points), confirming that \rQ\ gains come from retrieval rather than text conditioning.    However, providing RL-retrieved passages to the pre-RL model adds at most 3 EM points, whereas giving the RL agent control of \rR\ yields 11--22 EM points under both live and frozen retrieval.

\begin{table*}[t]
\centering\small
\setlength{\tabcolsep}{3pt}
\begin{tabular}{lllrrrccrrrr}
\toprule
recipient & type & data & pre-RL & RL & $\Delta$ & srch & valid & \rR & \rD & \rQ & \rA \\
\midrule
% Qwen3.5-2B & base & NQ & 4.6 & 13.0 & +8.4 & 0.6$\to$0.0 & 82$\to$95 & 22 & 73 & -8 & 13 \\
%  &  & HotpotQA & 4.9 & 13.0 & +8.1 & 0.6$\to$0.0 & 83$\to$98 & 7 & 98 & -8 & 3 \\
% Qwen3.5-4B & base & NQ & 12.9 & 26.0 & +13.1 & 1.9$\to$0.0 & 32$\to$100 & 22 & 45 & 0 & 32 \\
%  &  & HotpotQA & 8.2 & 23.0 & +14.8 & 2.1$\to$0.0 & 21$\to$100 & 35 & 53 & 2 & 10 \\
% Qwen3.5-2B & normal & NQ & 15.6 & 12.5 & -3.1 & 0.4$\to$0.0 & 94$\to$100 & \multicolumn{4}{c}{\scriptsize --- no gain ---} \\
%  &  & HotpotQA & 14.2 & 14.5 & +0.3 & 0.4$\to$0.0 & 92$\to$100 & \multicolumn{4}{c}{\scriptsize --- no gain ---} \\
% Qwen3.5-4B & normal & NQ & 22.8 & 19.2 & -3.6 & 1.2$\to$0.0 & 71$\to$100 & \multicolumn{4}{c}{\scriptsize --- no gain ---} \\
%  &  & HotpotQA & 22.3 & 20.6 & -1.7 & 2.0$\to$0.0 & 50$\to$100 & \multicolumn{4}{c}{\scriptsize --- no gain ---} \\
% \midrule
Qwen2.5-3B & base & NQ & 15.8 & 45.2 & +29.4 & 0.9$\to$1.1 & 53$\to$98 & \textbf{70} & 7 & 6 & 17 \\
 &  & HotpotQA & 9.2 & 38.9 & +29.7 & 0.9$\to$1.6 & 51$\to$97 & \textbf{81} & 1 & 5 & 13 \\
Qwen2.5-7B & base & NQ & 25.5 & 48.2 & +22.7 & 1.4$\to$1.1 & 79$\to$100 & \textbf{40} & 9 & 13 & 38 \\
 &  & HotpotQA & 18.3 & 44.1 & +25.8 & 1.6$\to$1.6 & 66$\to$100 & \textbf{45} & 18 & 5 & 32 \\
Qwen2.5-3B & instruct & NQ & 21.8 & 41.3 & +19.5 & 1.0$\to$1.3 & 97$\to$100 & 36 & 21 & 4 & \textbf{39} \\
 &  & HotpotQA & 22.0 & 34.9 & +12.9 & 1.2$\to$1.9 & 94$\to$99 & \textbf{69} & 12 & 4 & 15 \\
Qwen2.5-7B & instruct & NQ & 30.9 & 41.9 & +11.0 & 1.0$\to$1.2 & 97$\to$99 & 34 & 6 & 14 & \textbf{46} \\
 &  & HotpotQA & 30.7 & 39.0 & +8.3 & 1.3$\to$1.6 & 95$\to$99 & 35 & 6 & 14 & \textbf{45} \\
\midrule
Llama-3.2-1B & base & NQ & 2.1 & 17.2 & +15.1 & 0.2$\to$0.0 & 40$\to$100 & 27 & 26 & 0 & \textbf{47} \\
 &  & HotpotQA & 1.7 & 17.8 & +16.1 & 0.2$\to$0.0 & 18$\to$100 & \textbf{40} & 38 & 0 & 22 \\
Llama-3.2-3B & base & NQ & 12.6 & 44.2 & +31.6 & 0.7$\to$1.0 & 66$\to$100 & 33 & 18 & 15 & \textbf{34} \\
 &  & HotpotQA & 9.6 & 32.0 & +22.4 & 0.7$\to$1.0 & 63$\to$100 & 23 & 25 & 12 & \textbf{40} \\
Llama-3.2-1B & instruct & NQ & 12.7 & 28.8 & +16.1 & 0.8$\to$0.8 & 68$\to$98 & -23 & 41 & 16 & \textbf{66} \\
 &  & HotpotQA & 6.8 & 19.6 & +12.8 & 0.8$\to$0.8 & 57$\to$96 & 1 & 36 & 16 & \textbf{46} \\
Llama-3.2-3B & instruct & NQ & 29.0 & 32.2 & +3.2 & 1.4$\to$0.0 & 86$\to$100 & \multicolumn{4}{c}{\scriptsize --- no gain ---} \\
 &  & HotpotQA & 21.3 & 22.6 & +1.3 & 1.6$\to$0.0 & 83$\to$100 & \multicolumn{4}{c}{\scriptsize --- no gain ---} \\
\bottomrule
\end{tabular}
\caption{Main results showing the RL gain attribution study using our proposed \ourmethod framework. \emph{pre-RL}, \emph{RL}, and $\Delta$ refer to EM (\%) of the pre-RL agent, the RL agent, and their difference, where as \emph{srch} and \emph{valid} refer to searches per question and valid-answer rate (\%) shown as pre-RL agent $\to$ RL agent. Shares (\%) for roles (R: Reason, D: Decision, Q: Query, and A: Answer) are reported when the gain ($\Delta$) exceeds 4 points.}
\label{tab:grid_main}
\end{table*}

\section{Results}
\label{sec:results}

In this section, we present our main results in Table~\ref{tab:grid_main} and understand the findings across multiple factors, including question complexity, model scale, and instruction tuning.

\subsection{Effect of Question Complexity: Single-hop vs.\ Multi-hop}
\label{sec:res-dataset}

Question complexity has only a secondary effect on role distribution, mainly shifting credit between Reasoning and Answer while overall patterns remain driven by search behavior. Comparing role attributions across single-hop NQ and multi-hop HotpotQA for seven models (two models are excluded because their RL gain fell below the reporting threshold), we find that the Query role remains invariant across task types and stays bounded at or below 15\%. In contrast, Reasoning and Answer exhibit a clear trade-off: Reasoning attribution is consistently higher on multi-hop HotpotQA (across six of seven models with maximum up to 33 percentage points), whereas Answer attribution is higher on single-hop NQ (across six of seven models with maximum up to 29 percentage points), reflecting how multi-hop queries promote and shift credit from final answer generation toward intermediate reasoning steps. Meanwhile, the Decision share remains below 25\% when search behavior is stable (reaching non-zero search calls), but increases substantially when models either increase or decrease search calls (reaching up to 38--86\% when searches drop to zero). Overall, the distributional shift across question complexity is smaller than the divergence induced by changes in search behavior.

\subsection{Effect of Model Scale: Smaller vs.\ Larger Models}
\label{sec:res-size}

Model scale shifts which roles receive credit, moving attribution away from Reasoning and Decision toward Answer and Query. Across the three families where both sizes gain at least four points (Qwen2.5 base, Qwen2.5 instruct, and Llama-3.2 base), larger models allocate 17 to 40 percentage points more credit to Answer and Query combined in five of six comparisons. This shift is stronger in Qwen2.5 base, where the 3B model relies heavily on Reasoning (70\% vs.\ 40\% on NQ; 81\% vs.\ 45\% on HotpotQA) while the 7B model shifts 19 to 21 points more toward Answer; Qwen2.5 instruct mirrors this trade-off on HotpotQA (Reasoning 69\% vs.\ 35\%; Answer 15\% vs.\ 45\%). Similarly, in Llama-3.2 base, the 1B model abandons search (Query $\le 1\%$) whereas the 3B model sustains search calls and attributes 12--15\% to Query. Because larger models show higher pre-RL valid-answer rates (e.g., 53\% vs.\ 79\% on Qwen2.5 NQ) with the exception of Qwen2.5-Instruct, where rates are equal (97\%), the gain attributions generally suggest a scale-driven shift toward Answer and Query.

% preamble: \usepackage{colortbl}  \definecolor{mpos}{HTML}{649EE5}  \definecolor{mneg}{HTML}{EA7F7C}
% written by figs/interactions/make_table_moebius_grpo.py; do not edit by hand
\begin{table*}[t]
\centering\scriptsize
\setlength{\tabcolsep}{2.0pt}
\begin{tabular}{lr rrrr | rrrrrr rrrrr | rrrr}
\toprule
 & & \multicolumn{15}{c}{M\"obius terms} & \multicolumn{4}{c}{Shapley values} \\
\cmidrule(lr){3-17}\cmidrule(lr){18-21}
 & & \multicolumn{4}{c}{one role} & \multicolumn{6}{c}{two roles} & \multicolumn{5}{c}{three and four roles} & & & & \\
\cmidrule(lr){3-6}\cmidrule(lr){7-12}\cmidrule(lr){13-17}
agent & gain & \rR & \rD & \rQ & \rA & \rR$\times$\rD & \rR$\times$\rQ & \rR$\times$\rA & \rD$\times$\rQ & \rD$\times$\rA & \rQ$\times$\rA & \rR\rD\rQ & \rR\rD\rA & \rR\rQ\rA & \rD\rQ\rA & \rR\rD\rQ\rA & \rR & \rD & \rQ & \rA \\
\midrule
\multicolumn{21}{c}{\emph{NQ}} \\
\cmidrule(lr){1-21}
Qwen2.5-3B base & 29.4 & \cellcolor{mpos!100}+18.8 & \cellcolor{mpos!7}+0.6 & \cellcolor{mpos!28}+2.5 & \cellcolor{mpos!64}+5.7 & \cellcolor{mpos!22}+2.0 & \cellcolor{mneg!12}$-$1.1 & \cellcolor{mneg!25}$-$2.2 & \cellcolor{mneg!19}$-$1.7 & \cellcolor{mneg!33}$-$2.9 & \cellcolor{mneg!16}$-$1.4 & \cellcolor{mpos!20}+1.8 & \cellcolor{mpos!58}+5.2 & \cellcolor{mpos!18}+1.6 & \cellcolor{mpos!24}+2.1 & \cellcolor{mneg!18}$-$1.6 & \textbf{+20.6} & +1.9 & +1.8 & +5.0 \\
Qwen2.5-3B instruct & 19.5 & \cellcolor{mpos!100}+11.0 & \cellcolor{mpos!70}+6.2 & \cellcolor{mpos!40}+3.6 & \cellcolor{mpos!100}+13.8 & \cellcolor{mneg!28}$-$2.5 & \cellcolor{mneg!49}$-$4.4 & \cellcolor{mneg!100}$-$8.9 & \cellcolor{mneg!19}$-$1.7 & \cellcolor{mneg!82}$-$7.3 & \cellcolor{mneg!46}$-$4.1 & \cellcolor{mpos!33}+2.9 & \cellcolor{mpos!87}+7.7 & \cellcolor{mpos!46}+4.1 & \cellcolor{mpos!36}+3.2 & \cellcolor{mneg!46}$-$4.1 & +7.0 & +4.0 & +0.9 & \textbf{+7.6} \\
Qwen2.5-7B base & 22.7 & \cellcolor{mpos!100}+11.7 & \cellcolor{mpos!53}+4.7 & \cellcolor{mpos!40}+3.6 & \cellcolor{mpos!97}+8.6 & \cellcolor{mneg!52}$-$4.6 & \cellcolor{mneg!7}$-$0.6 & \cellcolor{mpos!10}+0.9 & \cellcolor{mneg!18}$-$1.6 & \cellcolor{mneg!8}$-$0.7 & \cellcolor{mpos!1}+0.1 & \cellcolor{mpos!24}+2.1 & \cellcolor{mpos!4}+0.4 & \cellcolor{mneg!12}$-$1.1 & \cellcolor{mpos!34}+3.0 & \cellcolor{mneg!43}$-$3.8 & \textbf{+9.1} & +2.1 & +2.9 & +8.6 \\
Llama-3.2-3B base & 31.6 & \cellcolor{mpos!1}+0.1 & \cellcolor{mneg!28}$-$2.5 & \cellcolor{mpos!27}+2.4 & \cellcolor{mpos!79}+7.0 & \cellcolor{mpos!99}+8.8 & \cellcolor{mpos!4}+0.4 & \cellcolor{mpos!24}+2.1 & \cellcolor{mneg!27}$-$2.4 & \cellcolor{mpos!8}+0.7 & \cellcolor{mneg!3}$-$0.3 & \cellcolor{mpos!79}+7.0 & \cellcolor{mpos!45}+4.0 & \cellcolor{mpos!6}+0.5 & \cellcolor{mpos!11}+1.0 & \cellcolor{mpos!31}+2.8 & +10.3 & +5.8 & +4.8 & \textbf{+10.8} \\
\midrule
\multicolumn{21}{c}{\emph{HotpotQA}} \\
\cmidrule(lr){1-21}
Qwen2.5-3B base & 29.7 & \cellcolor{mpos!100}+21.4 & \cellcolor{mneg!6}$-$0.5 & \cellcolor{mpos!17}+1.5 & \cellcolor{mpos!47}+4.2 & \cellcolor{mpos!31}+2.8 & \cellcolor{mpos!30}+2.7 & \cellcolor{mpos!9}+0.8 & \cellcolor{mneg!7}$-$0.6 & \cellcolor{mneg!9}$-$0.8 & \cellcolor{mneg!1}$-$0.1 & \cellcolor{mneg!7}$-$0.6 & \cellcolor{mpos!12}+1.1 & \cellcolor{mneg!20}$-$1.8 & \cellcolor{mpos!4}+0.4 & \cellcolor{mneg!9}$-$0.8 & \textbf{+23.9} & +0.3 & +1.6 & +3.9 \\
Qwen2.5-3B instruct & 12.9 & \cellcolor{mpos!100}+10.3 & \cellcolor{mpos!10}+0.9 & \cellcolor{mpos!29}+2.6 & \cellcolor{mpos!45}+4.0 & \cellcolor{mpos!11}+1.0 & \cellcolor{mneg!51}$-$4.5 & \cellcolor{mneg!51}$-$4.5 & \cellcolor{mneg!24}$-$2.1 & \cellcolor{mneg!3}$-$0.3 & \cellcolor{mneg!35}$-$3.1 & \cellcolor{mpos!37}+3.3 & \cellcolor{mpos!9}+0.8 & \cellcolor{mpos!56}+5.0 & \cellcolor{mpos!15}+1.3 & \cellcolor{mneg!20}$-$1.8 & \textbf{+8.9} & +1.6 & +0.5 & +2.0 \\
Qwen2.5-7B base & 25.8 & \cellcolor{mpos!100}+9.9 & \cellcolor{mpos!65}+5.8 & \cellcolor{mpos!21}+1.9 & \cellcolor{mpos!70}+6.2 & \cellcolor{mpos!15}+1.3 & \cellcolor{mneg!12}$-$1.1 & \cellcolor{mpos!88}+7.8 & \cellcolor{mneg!7}$-$0.6 & \cellcolor{mpos!21}+1.9 & \cellcolor{mpos!1}+0.1 & \cellcolor{mpos!7}+0.6 & \cellcolor{mneg!91}$-$8.1 & 0.0 & \cellcolor{mneg!8}$-$0.7 & \cellcolor{mpos!9}+0.8 & \textbf{+11.6} & +4.6 & +1.3 & +8.4 \\
Llama-3.2-3B base & 22.4 & \cellcolor{mneg!2}$-$0.2 & \cellcolor{mneg!6}$-$0.5 & \cellcolor{mpos!2}+0.2 & \cellcolor{mpos!57}+5.1 & \cellcolor{mpos!61}+5.4 & 0.0 & \cellcolor{mpos!25}+2.2 & \cellcolor{mpos!2}+0.2 & \cellcolor{mpos!34}+3.0 & \cellcolor{mpos!12}+1.1 & \cellcolor{mpos!34}+3.0 & \cellcolor{mneg!2}$-$0.2 & \cellcolor{mneg!3}$-$0.3 & \cellcolor{mpos!1}+0.1 & \cellcolor{mpos!37}+3.3 & +5.3 & +5.6 & +2.6 & \textbf{+8.9} \\
\bottomrule
\end{tabular}
\caption{M\"obius terms and Shapley values (EM points) of four GRPO agents on NQ and HotpotQA. Roles R, D, Q, and A refer to the same in Table~\ref{tab:grid_main}. The 15 M\"obius terms of a row sum to the gain. Cell shading highlights positive terms in blue and negative terms in red. Largest Shapley value is in bold.}
\label{tab:moebius_grpo}
\end{table*}

\subsection{Effect of Instruction Tuning: Base vs.\ Instruct}
\label{sec:res-instruct}

Instruction tuning does not produce a uniform shift in credit allocation; instead, its effect depends on how post-training alters a model's valid-answer rate and search trajectory. Across four base--instruct pairs, we find that instruction-tuned models with high valid-answer rates realize smaller RL gains and shift attribution toward Answer, whereas models that abandon search stall the RL gains and break the pattern. Specifically, in the Qwen2.5 family (3B and 7B), instruct variants start 18 to 44 percentage points higher in valid-answer rate than their base counterparts, leaving less format learning for RL and yielding 9.9 to 17.5 points lower overall gains while shifting credit from Reasoning ($-6$ to $-34$ percentage points) to Answer ($+8$ to $+22$ points on NQ). Conversely, Llama-3.2-1B reverses this trend: both variants begin with low valid-answer rates, yet the instruct model gains 6.1 to 11.2 points more because it preserves search (0.9 to 1.0 vs.\ 0.2 to 0.0 for base), raising Query attribution by 10 to 12 percentage points. Finally, Llama-3.2-3B instruct illustrates the combined effect of both behaviors—starting at a high valid-answer rate (86\% vs.\ 66\% on NQ) while abandoning search entirely (1.4--1.6 to 0.0), resulting in negligible gains ($+1.3$ to $+3.2$ points vs.\ $+22.4$ to $+31.6$ for base). Together, these comparisons demonstrate that pre-RL answer validity and search retention predict role attribution far more reliably than instruction tuning alone.

\section{Analysis}
\label{sec:analysis}

We examine four representative Search-R1 checkpoints in depth to uncover how roles interact and whether the RL optimizer alters credit attribution. Starting from Qwen2.5-3B base—which exhibits the largest Reasoning share in Table~\ref{tab:grid_main}—we select three controlled ablations that each vary a single axis: instruction tuning (Qwen2.5-3B instruct), model scale within the same family (Qwen2.5-7B base), and model family at comparable scale (Llama-3.2-3B base). Because public GRPO and PPO checkpoints exist for all four models, this unified selection supports both our role-interaction and optimizer-sensitivity analyses below.

\subsection{Role Interactions: Pairwise and Higher-Order Terms}
\label{sec:interactions}

While Shapley values quantify how much credit each role receives, M\"obius interaction terms (Eq.~\ref{eq:moebius}; Table~\ref{tab:moebius_grpo}) reveal whether those gains arise independently or through multi-role coordination. A pairwise interaction is positive when two roles produce synergistic gains beyond the sum of their individual effects, and negative when they overlap as substitutes. Across our four focal checkpoints and two datasets, pairwise terms already reshape how Table~\ref{tab:grid_main} shares should be read: in Llama-3.2-3B base, \rR$\times$\rD\ is the largest pairwise term on both NQ ($+8.8$) and HotpotQA ($+5.4$) while solo Reasoning ($+0.1$ and $-0.2$) and Decision ($-2.5$ and $-0.5$) contribute little or nothing on their own, showing that both roles must act together to improve exact match. Conversely, in Qwen2.5-3B instruct on NQ, negative pairwise terms with Answer (\rR$\times$\rA\ $-8.9$, \rD$\times$\rA\ $-7.3$) alongside strong solo contributions ($+11.0$ for Reasoning, $+6.2$ for Decision, and $+13.8$ for Answer against a $19.5$ total gain) indicate redundant overlap rather than joint synergy, while Qwen2.5-7B base shows task sensitivity (single-hop vs. multi-hop) as \rR$\times$\rA\ shifts from $+0.9$ on NQ to $+7.8$ on HotpotQA.

Higher-order M\"obius terms further clarify multi-role dynamics, provided that they are interpreted alongside lower-order effects rather than in isolation. Because a three- or four-way term isolates what remains after subtracting lower-order terms, a positive higher-order value can offset negative pairwise interactions rather than prove simultaneous synergy across all included roles as seen on NQ for \rR$\times$\rD$\times$\rA\ in Qwen2.5-3B base ($+5.2$ offsetting \rR$\times$\rA\ $-2.2$ and \rD$\times$\rA\ $-2.9$) and Qwen2.5-3B instruct ($+7.7$ offsetting $-8.9$ and $-7.3$).
% , or for \rR$\times$\rD$\times$\rA\ ($-8.1$) in Qwen2.5-7B base once the RL Decision policy is active. 
By contrast, Llama-3.2-3B base exhibits a positive \rR$\times$\rD$\times$\rQ\ term ($+7.0$ on NQ) alongside modest or non-negative pairwise terms (\rR$\times$\rD\ $+8.8$, \rR$\times$\rQ\ $+0.4$, \rD$\times$\rQ\ $-2.4$), indicating that RL Query adds incremental value once Reasoning and Decision are jointly engaged. Together, Shapley values identify where credit lands, while M\"obius terms show whether a share reflects standalone competence (Qwen2.5-3B base), coupled co-adaptation (Llama-3.2-3B base), or overlapping roles (Qwen2.5-3B instruct). This analysis can guide iterative RL training by indicating whether reward shaping should target individual roles or joint role combinations.

\subsection{RL Optimizer Sensitivity: GRPO vs.\ PPO}
\label{sec:optimiser}

\begin{table}[t]
\centering\footnotesize
\setlength{\tabcolsep}{1.9pt}
\begin{tabular}{lll|r|r|rrrr}
\toprule
model & data & opt. & $\Delta$ & srch & \rR & \rD & \rQ & \rA \\
\midrule
Qwen2.5-3B & NQ & GRPO & +29.4 & 1.1 & \textbf{70} & 7 & 6 & 17 \\
base &  & PPO & +27.7 & 2.1 & \textbf{83} & -1 & 2 & 16 \\
 & Hotpot & GRPO & +29.7 & 1.6 & \textbf{81} & 1 & 5 & 13 \\
 &  & PPO & +26.8 & 2.2 & \textbf{67} & 7 & 8 & 18 \\
\midrule
Qwen2.5-3B & NQ & GRPO & +19.5 & 1.3 & 36 & 21 & 4 & \textbf{39} \\
instruct &  & PPO & +12.8 & 1.9 & 31 & 17 & 6 & \textbf{46} \\
 & Hotpot & GRPO & +12.9 & 1.9 & \textbf{69} & 12 & 4 & 15 \\
 &  & PPO & +11.3 & 2.1 & \textbf{84} & 7 & -9 & 18 \\
\midrule
Qwen2.5-7B & NQ & GRPO & +22.7 & 1.1 & \textbf{40} & 9 & 13 & 38 \\
base &  & PPO & +20.8 & 4.0 & -27 & \textbf{77} & 11 & 39 \\
 & Hotpot & GRPO & +25.8 & 1.6 & \textbf{45} & 18 & 5 & 32 \\
 &  & PPO & +26.6 & 3.9 & 20 & \textbf{50} & 10 & 20 \\
\midrule
Llama-3.2-3B & NQ & GRPO & +31.6 & 1.0 & 33 & 18 & 15 & \textbf{34} \\
base &  & PPO & +28.1 & 1.0 & \textbf{38} & 17 & 13 & 32 \\
 & Hotpot & GRPO & +22.4 & 1.0 & 23 & 25 & 12 & \textbf{40} \\
 &  & PPO & +21.9 & 1.0 & 25 & 22 & 16 & \textbf{37} \\
\bottomrule
\end{tabular}
\caption{Role gain attribution between GRPO vs.\ PPO public checkpoints. $\Delta$, \emph{srch}, R, D, Q, and A refer to the same in Table~\ref{tab:grid_main}. Largest share among roles are in bold.}
\label{tab:optimiser}
\end{table}

The choice of RL optimizer alters the leading role only when it pushes search behavior to the turn limit; otherwise, the starting checkpoint governs which roles capture the performance gain. Comparing public GRPO and PPO checkpoints across four models (Table~\ref{tab:optimiser}), we find that the dominant roles remain consistent under both optimizers for three architectures: Reasoning leads in Qwen2.5-3B base (70\% under GRPO vs.\ 83\% under PPO on NQ); Qwen2.5-3B instruct favors Answer on NQ (39\% vs.\ 46\%) and Reasoning on HotpotQA (69\% vs.\ 84\%); and Llama-3.2-3B base balances Reasoning and Answer within six points on NQ (33\% and 34\% under GRPO vs.\ 38\% and 32\% under PPO) while favoring Answer on HotpotQA (40\% vs.\ 37\%). Notably, this invariance holds even when PPO increases search frequency over GRPO in both Qwen2.5-3B variants (2.1 vs.\ 1.1 and 1.9 vs.\ 1.3 searches per question on NQ), indicating that moderate shifts in search depth do not reorganize role attribution. The sole exception is Qwen2.5-7B base, where PPO saturates the four-search turn limit on NQ, shifting 77\% of credit to Decision and yielding a negative standalone Reasoning share ($-27$\%), whereas GRPO averages 1.1 searches and splits credit evenly between Reasoning (40\%) and Answer (38\%). Overall, the starting model rather than the RL optimizer establishes the role distribution, with the optimizer disrupting it only when driving search depth to the turn boundary.

\section{Related Work}
\label{sec:related}

Prior work on RL for search agents primarily evaluates end-to-end task performance rather than decomposing which behavioral changes produce the gain. Existing RL search agents \cite{jin2025searchr1,jin2025empirical,song2025r1searcher,chen2025research,zheng2025deepresearcher,sun2025zerosearch,gao2025asearcher} optimize outcome-based rewards and report aggregate Exact Match (EM) scores. Methods that restrict optimization to query generation \cite{jiang2025s3} or assign separate rewards to search and answer steps \cite{decoupling2025} pre-specify which trajectory roles require improvement rather than measuring their empirical contributions, while ablations on training configurations~\cite{searchr1pp2026} vary algorithmic hyperparameters rather than functional roles. Furthermore, intra-policy credit-assignment methods, including step-level rewards \cite{wang2025stepsearch,cwgrpo2026}, turn deletion \cite{lotapo2026}, counterfactual branching \cite{ricepo2026}, document replay \cite{bridgeevidence2026}, and per-role advantage estimation \cite{rdpo2026}, compute signals exclusively within a single policy without comparing the trained agent against its pre-RL initialization. \ourmethod directly measures the performance difference between the trained policy and its initialization across functional roles.

A second line of research combines multiple models during inference or localizes post-training behavioral changes. \citet{basemodelsknow2025} and \citet{newmodeloldhabits2026} switch between base and post-trained checkpoints at the individual token level during closed-book generation, whereas \ourmethod assigns complete functional roles across multi-turn trajectories with live retriever feedback to obtain an exact Shapley attribution of the RL gain. To localize RL improvements, \citet{frankenstein2026} attribute gains to parameter subsets in weight space, whereas \citet{sameanswer2026} demonstrate that out-of-domain gains from math RL with verifiable rewards stem largely from answer formatting, a pattern consistent with our Qwen2.5 findings. Finally, ShapleyFlow \cite{yang2025shapleyflow} applies Shapley values to static modules within a fixed agentic workflow, whereas \ourmethod attributes the training gain between two policies across functional trajectory roles during interactive retrieval.

\section{Conclusion}
\label{sec:conclusion}

We introduce \ourmethod, a framework that decomposes the EM gain of an RL-trained search agent across its four trajectory roles under closed-loop execution, showing that pre-RL answer validity, search retention, and model scale govern whether gains stem from reasoning--decision coupling, fixed task-level search budgets, or final answer extraction, while search query generation (\rQ) never exceeds 15\% of the total gain on NQ and HotpotQA. Because outcome-based EM obscures which functional behavior improves during training, future RL reward designs targeting retrieval skill should explicitly supervise query formulation (\rQ) and evidence-sufficient stopping decisions (\rD).

\section*{Limitations}

\ourmethod attributes the performance difference between a pair of policies rather than evaluating a standalone checkpoint, and its interpretations are subject to the properties of cooperative game-theoretic decomposition and our experimental setup. Attribution shares measure gains relative to a chosen baseline policy (the pre-RL initialization), and Shapley values average marginal contributions across all role permutations; when multi-role interactions are substantial (such as in the Qwen2.5-7B PPO pair), additive Shapley shares alone do not capture joint dependencies and must be interpreted alongside the M\"obius interaction terms. In addition, functional boundaries are segmented using structural action tags, grouping internal planning, passage synthesis, and structural formatting together inside the intermediate reasoning role (\rR), leaving finer-grained decompositions of \rR\ to future work.

Our empirical evaluation is also bounded by model scale, inference serving stochasticity, and diagnostic scope. Compute constraints limit our locally trained models to 7B parameters or smaller, requiring us to omit a 14B or 32B model pair, and batched greedy decoding introduces approximately one EM point of serving variance. Furthermore, the fixed-budget and retrieval-swap diagnostics cover only the two Qwen2.5 PPO pairs on NQ and HotpotQA, and selecting the optimal budget on the evaluation set provides an upper-bound estimate within this noise floor. Finally, training recipes incorporating query-level process rewards, longer optimization schedules, or live web search may redistribute gains differently across roles, which \ourmethod can be directly applied to measure.

\bibliography{refs}

\clearpage
\appendix
% ---------------------------------------------------------------------------------------------------------------------
% Appendix. Needs \usepackage{xcolor}, graphicx, booktabs, amsmath and the role macros \rR \rD \rQ \rA of main.tex.
% Generated parts: tables/tab_shapley_from_moebius.tex and appendix_examples.tex (appendix_gen/make_appendix_parts.py),
% figs/interactions/fig_matrix_grpo4.pdf (figs/interactions/make_fig_matrix_grpo4.py).
% ---------------------------------------------------------------------------------------------------------------------
\definecolor{roleR}{HTML}{2A78D6}\definecolor{roleD}{HTML}{EB6834}\definecolor{roleQ}{HTML}{1BAF7A}
\definecolor{roleA}{HTML}{EDA100}\definecolor{roleE}{HTML}{C9C8C2}
\definecolor{roleRt}{HTML}{1F5FB0}\definecolor{roleDt}{HTML}{B4420F}\definecolor{roleQt}{HTML}{0E7A54}
\definecolor{roleAt}{HTML}{8A5A00}\definecolor{roleEt}{HTML}{5F5E5A}
\newcommand{\chip}[3]{{\setlength{\fboxsep}{1.2pt}\colorbox{#1}{\textcolor{#2}{\textsf{\bfseries\scriptsize #3}}}}}
\newcommand{\spR}[1]{\chip{roleR}{white}{R}\,\textcolor{roleRt}{#1}}
\newcommand{\spD}[1]{\chip{roleD}{white}{D}\,\textcolor{roleDt}{#1}}
\newcommand{\spQ}[1]{\chip{roleQ}{black}{Q}\,\textcolor{roleQt}{#1}}
\newcommand{\spA}[1]{\chip{roleA}{black}{A}\,\textcolor{roleAt}{#1}}
\newcommand{\spE}[1]{\chip{roleE}{black}{env}\,\textcolor{roleEt}{\emph{#1}}}

\section{From M\"obius Terms to Shapley Values}
\label{app:moebius}
M\"obius interaction terms and Shapley values represent two complementary decompositions of the same 16 combination measurements.
For each subset $S \subseteq N$, \ourmethod measures $v(S)$, the EM score obtained when the RL-trained agent owns the roles in $S$ and the pre-RL initialization owns the remaining roles $N \setminus S$.
The M\"obius inversion partitions the total gain into 15 components, one for each non-empty subset $T \subseteq N$: $m(T)$ captures the marginal synergy that emerges only when the RL agent owns all roles in $T$ simultaneously, after subtracting all lower-order effects explained by subsets of $T$ (Eq.~\ref{eq:moebius}).
Consequently, the gain of any combination $S$ equals the sum of the M\"obius terms composed within it:
\begin{equation}
v(S)-v(\emptyset)=\sum_{T \subseteq S,\, T \neq \emptyset} m(T),
\label{eq:moebius_sum}
\end{equation}
and for the grand combination $S=N$, all 15 terms sum to the total RL gain $v(N) - v(\emptyset)$.

A Shapley value assigns each role its fair share of these interaction components by dividing every term $m(T)$ equally among its participating roles:
\begin{equation}
\phi_i=\sum_{T\ni i}\frac{m(T)}{|T|}.
\label{eq:shapley_moebius}
\end{equation}
For intermediate reasoning (\rR), $\phi_\rR$ equals the standalone term $m(\{\rR\})$, plus half of the three pairwise terms containing \rR, one-third of the three three-way terms containing \rR, and one-quarter of the four-way term.
Thus, M\"obius terms identify which roles (individually or jointly) generate performance gains, whereas Shapley values summarize each role's overall contribution into a single scalar without retaining higher-order coupling structure.

% written by appendix_gen/make_appendix_parts.py; do not edit by hand
\begin{table}[!ht]
\centering\footnotesize
\setlength{\tabcolsep}{3.5pt}
\begin{tabular}{lrrrrr}
\toprule
role & alone & $\tfrac12$ pairs & $\tfrac13$ triples & $\tfrac14$ four & Shapley \\
\midrule
\multicolumn{6}{l}{\emph{Qwen2.5-3B base, NQ (gain 29.4)}} \\
\rR & +18.8 & $-$0.7 & +2.9 & $-$0.4 & +20.6 \\
\rD & +0.6 & $-$1.3 & +3.0 & $-$0.4 & +1.9 \\
\rQ & +2.5 & $-$2.1 & +1.8 & $-$0.4 & +1.8 \\
\rA & +5.7 & $-$3.3 & +3.0 & $-$0.4 & +5.0 \\
\midrule
\multicolumn{6}{l}{\emph{Llama-3.2-3B base, NQ (gain 31.6)}} \\
\rR & +0.1 & +5.7 & +3.8 & +0.7 & +10.3 \\
\rD & $-$2.5 & +3.6 & +4.0 & +0.7 & +5.8 \\
\rQ & +2.4 & $-$1.2 & +2.8 & +0.7 & +4.8 \\
\rA & +7.0 & +1.3 & +1.8 & +0.7 & +10.8 \\
\bottomrule
\end{tabular}
\caption{Shapley values rebuilt from the M\"obius terms of Table~\ref{tab:moebius_grpo} (EM points). \emph{alone}: the single-role term; the next three columns: half of the sum of the two-role terms that contain the role, a third of the three-role terms and a quarter of the four-role term. The four parts add up to the Shapley value, up to rounding.}
\label{tab:shapley_from_moebius}
\end{table}

Table~\ref{tab:shapley_from_moebius} illustrates Eq.~\ref{eq:shapley_moebius} on NQ across two contrasting checkpoints.
For Qwen2.5-3B base, higher-order interaction terms are negligible: standalone reasoning achieves $+18.8$ EM points and its Shapley value is $20.6$, indicating that its attribution closely reflects independent reasoning improvements.
For Llama-3.2-3B base, the opposite pattern holds: standalone reasoning gains only $+0.1$ points, yet its Shapley value reaches $10.3$, driven by $5.7$ points from pairwise interactions (primarily \rR$\times$\rD\ at $+8.8$) and $4.5$ points from three- and four-way terms.
Because this attribution arises only when the RL policy simultaneously controls decision-making, $\phi_\rR$ cannot be interpreted independently of $\phi_\rD$.

% \paragraph{Interpreting Higher-Order Terms.}
% A pairwise term is positive when two roles achieve greater gains together than the sum of their individual effects (synergy), negative when either role alone already captures part of the gain (redundancy), and zero when their effects are purely additive.
% Three- and four-way terms cannot be interpreted in isolation because they represent residuals after subtracting all embedded pairwise terms.
% Summing a higher-order term with its constituent pairwise terms yields, via Eq.~\ref{eq:moebius_sum}, the net non-additive gain of the subset beyond its standalone role effects, whose sign carries the same interpretation as in the pairwise case.
% For Qwen2.5-3B-Instruct on NQ, \rR$\times$\rD$\times$\rA is $+7.7$, but combining it with its three embedded pairwise terms ($-2.5$, $-8.9$, and $-7.3$) yields $-11.0$, indicating net overlap among the three roles.
% Conversely, for Llama-3.2-3B base on NQ, \rR$\times$\rD$\times$\rQ is $+7.0$, and summing it with its pairwise terms ($+8.8$, $+0.4$, and $-2.4$) yields $+13.8$, confirming that all three roles are required jointly.

% declared here so that it is placed at the top of the page on which the next section is discussed
\begin{figure*}[t]
\centering
\includegraphics[width=\textwidth]{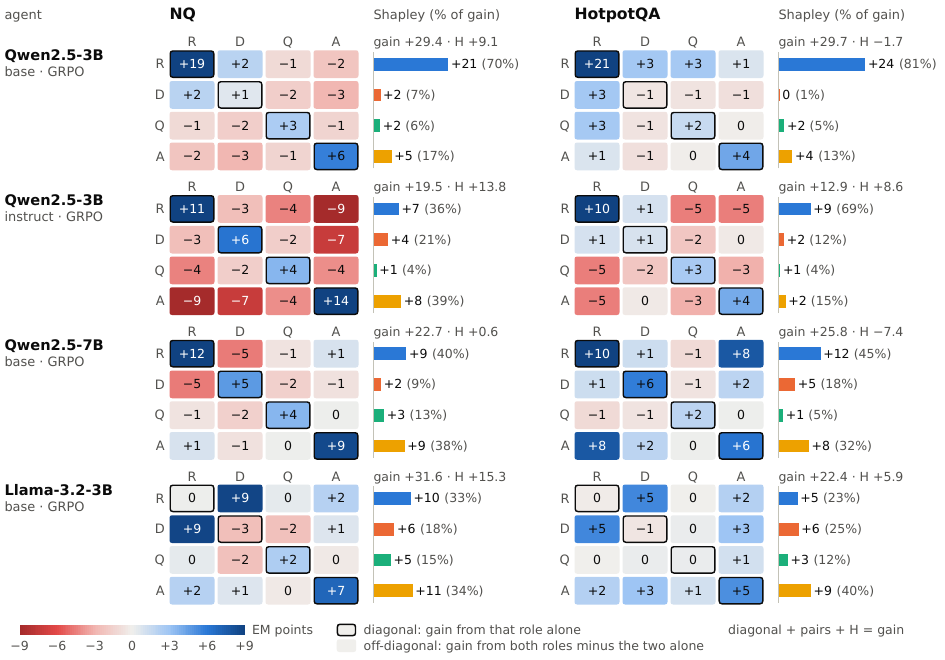}
\caption{Interaction matrices for the four GRPO-trained agents in Table~\ref{tab:moebius_grpo} on NQ (left) and HotpotQA (right), measured in EM points. Outlined diagonal entries denote standalone role gains when the RL agent controls a single role. Off-diagonal entries show pairwise M\"obius terms (joint two-role gain minus the sum of standalone gains). Right-hand bars report each role's Shapley value in EM points (and percentage share of total gain). $H$ denotes the unweighted sum of all three- and four-role M\"obius terms, whereas diagonal entries, upper-triangular pairs, and $H$ sum to the total RL gain. Color shading is shown in the heatmap bar that saturates at $\pm 9$ points.}
\label{fig:matrix_grpo4}
\end{figure*}

\section{Interaction Matrices}
\label{app:matrices}
Figure~\ref{fig:matrix_grpo4} visualizes the standalone and pairwise M\"obius terms from Table~\ref{tab:moebius_grpo} across agents and datasets.
Diagonal entries display single-role gains, off-diagonal entries display pairwise interactions, right-hand bars show Shapley values, and $H$ aggregates the three- and four-role terms omitted from the 2D grid.
Three structural regimes emerge:
(1) For Qwen2.5-3B base, the standalone reasoning diagonal ($+19$ on NQ, $+21$ on HotpotQA) dominates all other entries, while every pairwise interaction remains within $\pm 3$ points, confirming largely additive role contributions.
(2) For Qwen2.5-3B-Instruct on NQ, all four diagonal terms are positive ($+4$ to $+14$) whereas all six pairwise terms are negative ($-2$ to $-9$), demonstrating substantial cross-role redundancy.
(3) For Llama-3.2-3B base, the standalone diagonals for reasoning and decision are zero or negative ($0$ and $-3$ on NQ; $0$ and $-1$ on HotpotQA), while \rR$\times$\rD forms the largest pairwise entry on both datasets ($+9$ and $+5$), showing strong coupling between reasoning and search termination.
Finally, $H$ is substantial for Qwen2.5-3B-Instruct ($+13.8$) and Llama-3.2-3B base ($+15.3$) on NQ, requiring joint inspection of the matrix alongside the higher-order columns of Table~\ref{tab:moebius_grpo}.

\section{Example Trajectories}
\label{app:examples}
We present qualitative case studies illustrating three distinct attribution patterns: gains driven by intermediate reasoning (\rR), final answer extraction (\rA), and search-or-answer decisions (\rD).
For each case, we select an evaluation instance where the pre-RL initialization fails, the RL-trained agent succeeds, and performance across the 16 role assignments tracks control of the target role.
Figures~\ref{fig:example_r}--\ref{fig:example_d} display the complete trajectories (with line breaks collapsed and retrieved passage text omitted for readability) with color-coding each generated span by the active role policy.
% the three example boxes are declared here and allowed to share the top of the next page
\setcounter{dbltopnumber}{3}\setcounter{totalnumber}{4}\renewcommand{\dbltopfraction}{0.95}\renewcommand{\textfraction}{0.03}
% written by appendix_gen/make_appendix_parts.py; do not edit by hand
\begin{figure*}[t]
\scriptsize
\textbf{Question:} where was the first sonic drive in located?\hfill\textbf{Gold answer:} Woodward, Oklahoma\par\vspace{2pt}\hrule\vspace{4pt}
\begin{minipage}[t]{0.485\textwidth}\raggedright
\textbf{Pre-RL agent alone} (answer ``San Bernardino, California'', EM 0, 1 search)\par\vspace{2pt}
\textbf{Turn 1.} \spR{I will conduct reasoning and search for the answer to the question. \texttt{<think>} The first Sonic Drive-In was founded in 1969 by Don Thompson and his wife, Mary. It was located in San Bernardino, California. \texttt{</think>}} \spD{\texttt{<search>}} \spQ{first sonic drive in location \texttt{</search>}} \spE{3 passages returned}\par
\textbf{Turn 2.} \spD{\texttt{<answer>}} \spA{San Bernardino, California \texttt{</answer>}}\par
\end{minipage}\hfill
\begin{minipage}[t]{0.485\textwidth}\raggedright
\textbf{RL agent alone} (answer ``Woodward, Oklahoma'', EM 1, 1 search)\par\vspace{2pt}
\textbf{Turn 1.} \spR{\texttt{<think>} To answer this question, I need to find information about the first Sonic Drive-In location. This involves knowing the history of Sonic Drive-In and its founding location. \texttt{</think>}} \spD{\texttt{<search>}} \spQ{Where was the first Sonic Drive-In located \texttt{</search>}} \spE{3 passages returned}\par
\textbf{Turn 2.} \spR{\texttt{<think>} From the information provided, I found that the first Sonic Drive-In was located in Woodward, Oklahoma. \texttt{</think>}} \spD{\texttt{<answer>}} \spA{Woodward, Oklahoma \texttt{</answer>}}\par
\end{minipage}
\par\vspace{4pt}\hrule\vspace{3pt}
{\scriptsize\chip{roleR}{white}{R}\,reasoning\quad\chip{roleD}{white}{D}\,decision\quad\chip{roleQ}{black}{Q}\,query\quad\chip{roleA}{black}{A}\,answer\quad\chip{roleE}{black}{env}\,written by the environment}
\caption{Reasoning-led gain: Qwen2.5-3B base (GRPO) on NQ. The reasoning role receives 70\% of this agent's gain on NQ (Table~\ref{tab:grid_main}). Both models receive the same three passages, which contain the gold answer. EM on this question is 1 in all eight role assignments in which the RL agent writes \rR\ and 0 in the other eight.}
\label{fig:example_r}
\end{figure*}

\begin{figure*}[t]
\scriptsize
\textbf{Question:} what kind of beer is st pauli girl?\hfill\textbf{Gold answer:} Special Dark / Lager / Non-Alcoholic Malt Beverage\par\vspace{2pt}\hrule\vspace{4pt}
\begin{minipage}[t]{0.485\textwidth}\raggedright
\textbf{Pre-RL agent alone} (answer ``St. Pauli Girl is a German beer brand.'', EM 0, 1 search)\par\vspace{2pt}
\textbf{Turn 1.} \spR{\texttt{<think>} St. Pauli Girl is a popular beer brand from Germany. To find out what kind of beer it is, I will search for information about St. Pauli Girl. \texttt{</think>}} \spD{\texttt{<search>}} \spQ{St. Pauli Girl beer type \texttt{</search>}} \spE{3 passages returned}\par
\textbf{Turn 2.} \spD{\texttt{<answer>}} \spA{St. Pauli Girl is a German beer brand. \texttt{</answer>}}\par
\end{minipage}\hfill
\begin{minipage}[t]{0.485\textwidth}\raggedright
\textbf{RL agent alone} (answer ``Lager'', EM 1, 1 search)\par\vspace{2pt}
\textbf{Turn 1.} \spR{\texttt{<think>} To answer this question, I need to find out what kind of beer 'St. Pauli Girl' is. This involves identifying the type of beer it represents. \texttt{</think>}} \spD{\texttt{<search>}} \spQ{What kind of beer is St. Pauli Girl? \texttt{</search>}} \spE{3 passages returned}\par
\textbf{Turn 2.} \spR{\texttt{<think>} From the information provided, I found that 'St. Pauli Girl' is a beer. \texttt{</think>}} \spD{\texttt{<answer>}} \spA{Lager \texttt{</answer>}}\par
\end{minipage}
\par\vspace{4pt}\hrule\vspace{3pt}
{\scriptsize\chip{roleR}{white}{R}\,reasoning\quad\chip{roleD}{white}{D}\,decision\quad\chip{roleQ}{black}{Q}\,query\quad\chip{roleA}{black}{A}\,answer\quad\chip{roleE}{black}{env}\,written by the environment}
\caption{Answer-led gain: Qwen2.5-3B base (GRPO) on NQ. The answer role receives 17\% of this agent's gain on NQ (Table~\ref{tab:grid_main}). Both models receive the same three passages, which contain the gold answer. EM on this question is 1 in all eight role assignments in which the RL agent writes \rA\ and 0 in the other eight, and no answer written by the pre-RL model contains the gold answer.}
\label{fig:example_a}
\end{figure*}

\begin{figure*}[t]
\scriptsize
\textbf{Question:} What did the person in which the Toronto Pearson International Airport is named in honour of organize to win the Nobel Peace Prize?\hfill\textbf{Gold answer:} United Nations Emergency Force\par\vspace{2pt}\hrule\vspace{4pt}
\begin{minipage}[t]{0.485\textwidth}\raggedright
\textbf{Pre-RL agent alone} (answer ``Sir William Randal Cremer'', EM 0, 1 search)\par\vspace{2pt}
\textbf{Turn 1.} \spR{\texttt{<think>} The Toronto Pearson International Airport is named after Sir Samuel Leonard Tilley, who was a Canadian politician and businessman. To find out what he organized to win the Nobel Peace Prize, I need to search for information about his Nobel Peace Prize. \texttt{</think>}} \spD{\texttt{<search>}} \spQ{query: Sir Samuel Leonard Tilley Nobel Peace Prize \texttt{</search>}} \spE{3 passages returned}\par
\textbf{Turn 2.} \spR{\texttt{<think>} From the search results, I found that Sir William Randal Cremer was the one who organized to win the Nobel Peace Prize. He was awarded the Nobel Peace Prize in 1903 for his work with the international arbitration movement. \texttt{</think>}} \spD{\texttt{<answer>}} \spA{Sir William Randal Cremer \texttt{</answer>}}\par
\end{minipage}\hfill
\begin{minipage}[t]{0.485\textwidth}\raggedright
\textbf{RL agent alone} (answer ``the United Nations Emergency Force'', EM 1, 4 searches)\par\vspace{2pt}
\textbf{Turn 1.} \spR{\texttt{<think>} 1. I need to find out who the Toronto Pearson International Airport is named after. 2. I need to find out what this person organized to win the Nobel Peace Prize. \texttt{</think>}} \spD{\texttt{<search>}} \spQ{Toronto Pearson International Airport \texttt{</search>}} \spE{3 passages returned}\par
\textbf{Turn 2.} \spR{\texttt{<think>} 1. The Toronto Pearson International Airport is named after Lester B. Pearson. 2. I need to find out what Lester B. Pearson organized to win the Nobel Peace Prize. \texttt{</think>}} \spD{\texttt{<search>}} \spQ{Lester B. Pearson Nobel Peace Prize \texttt{</search>}} \spE{3 passages returned}\par
\textbf{Turn 3.} \spR{\texttt{<think>} 1. Lester B. Pearson won the Nobel Peace Prize. 2. I need to find out what he organized to win the prize. \texttt{</think>}} \spD{\texttt{<search>}} \spQ{Lester B. Pearson Suez Canal Crisis \texttt{</search>}} \spE{3 passages returned}\par
\textbf{Turn 4.} \spR{\texttt{<think>} 1. Lester B. Pearson won the Nobel Peace Prize for his work. 2. I need to find out what he organized to win the prize. \texttt{</think>}} \spD{\texttt{<search>}} \spQ{Lester B. Pearson organized to win the Nobel Peace Prize \texttt{</search>}} \spE{3 passages returned}\par
\textbf{Turn 5.} \spR{\texttt{<think>} 1. Lester B. Pearson won the Nobel Peace Prize in 1957. 2. I need to find out what he organized to win the prize. \texttt{</think>}} \spD{\texttt{<answer>}} \spA{the United Nations Emergency Force \texttt{</answer>}}\par
\end{minipage}
\par\vspace{4pt}\hrule\vspace{3pt}
{\scriptsize\chip{roleR}{white}{R}\,reasoning\quad\chip{roleD}{white}{D}\,decision\quad\chip{roleQ}{black}{Q}\,query\quad\chip{roleA}{black}{A}\,answer\quad\chip{roleE}{black}{env}\,written by the environment}
\caption{Decision-led gain: Qwen2.5-7B base (PPO) on HotpotQA. The decision role receives 50\% of this agent's gain on HotpotQA (\S\ref{sec:optimiser}). EM on this question is 1 in four of the eight role assignments in which the RL agent writes \rD\ and 0 in all eight in which the pre-RL model writes it.}
\label{fig:example_d}
\end{figure*}

\end{document}